\documentclass[doublecol]{epl2} 

\usepackage{amsmath,amssymb,mathrsfs}

\newcommand{\bk}{{\bf k}}

\newcommand{\bq}{\boldsymbol{q}}

\newcommand{\IC}{I_{\rm C}}
\newcommand{\IP}{I_{\rm P}}
\newcommand{\ka}{k_{1b}}
\newcommand{\kb}{k_{2b}}
\newcommand{\kc}{k_{3b}}
\newcommand{\kp}{k_b}
\newcommand{\Qp}{Q_b}
\newcommand{\qcp}{q_b}

\newcommand{\Hc}{H_{\rm c2}}
\newcommand{\HP}{H_{\rm P}}

\newcommand{\Tc}{T_{\rm c}}

\title{
On the Origin of the Anomalous Upper Critical Field in Quasi-One-Dimensional Superconductors}
\shorttitle{
On the Origin of the Anomalous $H_{\rm c2}$ in Quasi-1D Superconductors
} 

\author{Y. Fuseya\inst{1,2} \and C. Bourbonnais\inst{2,}\inst{3} \and K. Miyake\inst{1}}
\shortauthor{Y. Fuseya \etal}

\institute{                    
  \inst{1} Department of Materials Engineering Science, Osaka University, 1-3 Machikaneyama, Toyonaka, Osaka 560-8531, Japan\\
  \inst{2} Regroupement Qu\'ebecois sur les Mat\'eriaux de Pointe,
  D\'epartement de physique, Universit\'e de Sherbrooke, Sherbrooke,
  Qu\'ebec, Canada, J1K-2R1\\
\inst{3} Canadian Institute for Advanced Research, Toronto, Canada}

\abstract{
Upper critical field, $\Hc$,  in quasi-1D superconductors is investigated by the weak coupling renormalization group technique. 
	It is shown that $\Hc $ greatly exceeds not only the Pauli limit, but also the conventional paramagnetic limit of the Flude-Ferrell-Larkin-Ovchinnikov (FFLO) state.
	This increase is mainly due to  quasi-1D  fluctuations effect as triggered by interference between unconventional superconductivity and density-wave instabilities.
	Our results give a novel viewpoint on the large $\Hc$ observed in TMTSF-salts in terms of a $d$-wave FFLO state that is predicted to be verified by the $\Hc$ measurements under pressure.
	}

\begin{document}

\maketitle

\section{Introduction}
	
	Thirty years after the discovery of organic superconductivity in the quasi-one-dimensional (quasi-1D) molecular conductor (TMTSF)$_2$PF$_6$  \cite{Jerome1980}, the nature of this phase has not been entirely clarified in spite of   intensive experimental and theoretical efforts devoted to this end. An enduring problem concerns  the properties of superconductivity under magnetic field, in particular the issue of the upper critical field  $\Hc$  in these materials,  which is the topic of the present letter.
%
	%
	%
	%
	 The value of $\Hc$ determined for instance by  resistivity measurements, drastically exceeds the so-called Clogston or Pauli  limit, $\HP$. The excess occurs when the magnetic field is oriented parallel to the plane of (TMTSF)$_2 X$ ($X=$PF$_6$ \cite{IJLee1997}, ClO$_4$ \cite{Oh2004}), that is when pair breaking orbital effect is virtually quenched by anisotropy.
	There are basically two possibilities that have been put forward to explain the excess of $\Hc$:   the existence of a spin-triplet pairing or the presence of a  Fulde-Ferrell-Larkin-Ovchinnikov (FFLO) phase. 

	The constant spin susceptibility, $\chi$, as obtained from the first Knight shift measurements across $T_c$ for $X=$PF$_6$ under pressure gave support to the former scenario for triplet superconductivity\cite{IJLee2002}.
	On the other hand, more recent NMR  results on the ambient pressure superconductor $X=$ClO$_4$,  display a decreasing  $\chi$  below $\Tc$, at least in low fields, which is compatible with singlet rather than triplet pairing\cite{Shinagawa2007}.
	As for the FFLO state, an estimation of the  paramagnetic limit, $\Hc^{\rm FFLO}$, puts it  lower than the $\Hc$ observed in (TMTSF)$_2 X$ \cite{Lebed1999}, which would make this scenario unlikely on a theoretical basis. 	%
	%
	However, recent detailed resistivity measurements on $X=$ClO$_4$ compound using   rotating field,
	%
	display the  signature of   non-uniform superconductivity at high fields, suggesting that a  FFLO state could be realized\cite{Yonezawa2008a,Yonezawa2008b}.
	The amplitude of  $\Hc$  determined by the bulk measurements such as   the nuclear relaxation rate\cite{Shinagawa2007}, magneto-torque \cite{Satsukawa2009} and specific heat \cite{Yonezawa2012}, is lower than that of resistivity measurements, indicating the existence of a high field interval $\Hc^{\rm bulk}<H<\Hc^{\rm resist}$, which we term as   a ``transient region",  where    incomplete superconductivity is realized.
	%
	%
	
	On theoretical side, the possibility of triplet superconductivity in an interacting quasi-1D electron system has been investigated on a microscopic basis in various situations.
	In zero field, $f$-wave triplet pairing has been shown to compete with  $d$-wave singlet pairing, when long-range intrachain interactions become sufficiently large \cite{Kuroki2001,Fuseya2002,Tanaka2004,Fuseya2005}.
	The competition was also shown to emerge when interchain repulsive  interactions  are finite, albeit small \cite{Nickel2005,Nickel2006}; or as the amplitude of on-site interaction is huge \cite{Ohta2005}.
	%
	 A field-induced singlet to triplet pairing crossover has been also proposed  to occur on  phenomenological grounds  \cite{Fuseya2002,Fuseya2005,Shimahara2000}. 
	 A similar transition under field has been shown to take place microscopically using mean-field theory \cite{Belmechri0908}, or when in a similar framework, long-range interactions are included \cite{Aizawa2008,Aizawa2009,Kajiwara2009}.
	The $H$-$T$ phase diagram thus obtained, however, departs from observation obtained for the bulk\cite{Yonezawa2008a,Yonezawa2008b,Yonezawa2012}; 
	it would also give a temperature independent spin susceptibility in the metallic state, at variance with experiments \cite{Dumm2000,Fuseya2007a}.
	%
	%
	 
	The FFLO state in the quasi-1D geometry has  been examined in great details in the framework of the BCS mean field theory, which  predicts its existence in the region of small $T_c$\cite{Lebed1986,Dupuis1993,Lebed2011}.
	However, mean-field theory  neglects  the influence of  fluctuations, in particular, those linked to many-body processes resulting from interference  between superconducting (SC) and density-wave (DW) pairings. 
	Such a mix of pairing mechanisms is present at every order of perturbation theory for quasi-1D interacting  electron  systems and is responsible for the dynamical generation of $d$-wave superconducting pairing  from spin fluctuations\cite{Fuseya2007a,Duprat2001,Nickel2005,Bourbonnais2009}.  
	For compounds like (TMTSF)$_2 X$, where superconductivity is just found in the close proximity of a SDW instability, the interference  is expected to  play an  important role. 
	Fully taken into account by the renormalization group method (RG), the interference proved to have a non trivial impact not only on the mechanism and nature of the superconductivity\cite{Nickel2005,Nickel2006,Fuseya2005,Fuseya2007b}, but also on the property of the metallic state above $\Tc$. 
	It carries over  well below the crossover  temperature scale for the onset of transverse coherent single electron motion where marked deviations from an ordinary Fermi liquid can be found \cite{Fuseya2007a,Bourbonnais2009,Sedeki2010,DoironLeyraud2009}. 
	The repercussions of these fluctuation effects on the stability of superconductivity under field  are  until now unexplored.

	%
	%
	 %
	 %
	 %
	 %
	%
	
	In this letter  we tackle the  problem of upper critical field in quasi-1D superconductors close to a spin density-wave instability and clarify the influence of quasi-1D fluctuations on superconductivity under magnetic field.
	For this purpose, we extend the renormalization group ($N$-chains RG) technique \cite{Fuseya2007a} to incorporate a finite Zeeman splitting, and study the  influence of interference on $d$-wave singlet superconductivity under magnetic field. 	
	%
	%
	It is shown that the $\Hc$ exceed not only $\HP$, but also the mean-field prediction of $\Hc^{\rm FFLO}$ to a great extent. 
	The predicted  non universal  character of the effect, as a function of interactions and nesting alterations of the Fermi surface, correlates the degree of the interference between SC and DW.
	This  give the possibility to confirm the relevance of the present theory by new experiments under pressure and field.   
	%
	%

\section{Theory}
	We consider a quasi-1D model of a square array of  weakly coupling metallic chains, taken away from half-filling for simplicity.
	The hopping integrals $t_i$ obey the following anisotropic sequence $t_a \gg t_b \gg t_c$ along the $a$-, $b$-, and $c$-axis. 
	The magnetic field is oriented such that its impact is restricted to the Zeeman splitting alone.	
	Such conditions  can be realized in  practice 
	%
	when the field is oriented in the $ab$ plane, which easily suppresses the orbital motion along the least conducting $c$ axis at small enough $t_{c}$\cite{Lebed1986}.
	The remaining relevant electron spectrum in the remaining $ab$ plane is given by   
	\begin{align}
	E_{\pm}(\bk) = v_F(\pm k_a-k_F) + \xi_b (k_b) + \mu_B\sigma H,
	\end{align}
	  where $\mu_{\rm B}$ is the Bohr magneton (in the following, we use $h=\mu_{\rm B} H$ and  $\hbar=1$,  $k_B=1$), and $\xi_b(k_b)=-2t_b \cos k_b -2t_b' \cos 2k_b$.
	Here the longitudinal part of the spectrum has been linearized around $\pm k_F$, with the longitudinal Fermi velocity $v_F=2t_a\sin k_F$.
	The next-nearest interchain hopping, $t_b'$, breaks the perfect nesting condition of the Fermi surface (FS).

	Electron-electron  scattering matrix elements are expressed  in terms of interactions between electrons on opposite Fermi sheets \cite{Solyom1979}, namely  as $g_\| $ and  $g_{1\perp}$ for backward scattering with parallel and antiparallel spins, and $g_{2\perp}$ for forward scattering. 
	For the Hubbard model, we have  $g_{||}=0$, $g_{1\perp}=g_{2\perp}=U/\pi v_F\equiv \tilde{U}$, as bare  normalized couplings,   where $U$ is the on-site repulsion of the Hubbard model. 
	%
	%
	We employ a momentum shell RG approach that is essentially equivalent to the one-particle irreducible scheme at one-loop level\cite{Fuseya2007a,Bourbonnais2004,Nickel2006,Honerkamp2001}.

	The one-loop flow equations so obtained are
	%
\begin{align}
	\partial_\ell g_{||}&= 
	g_{||} g_{||}
	\left( I_{\rm C}^{0} -I_{\rm P}^{0}
	\right)
	%
	-
	g_{1\perp} g_{1\perp}
	I_{\rm P}^{4h}
	,
	\label{g||}\\
	\partial_\ell g_{1\perp}&=
	-
	\left(
	g_{1\perp} g_{2\perp}
	+g_{2\perp} g_{1\perp}
	\right)
	\left( I_{\rm C}^{0} 
	+I_{\rm C}^{4h}  \right)/2
	\nonumber\\&
	-
	\left(
	g_{1\perp} g_{||}
	+g_{||} g_{1\perp}
	\right)
	\left( I_{\rm P}^{0} 
	+I_{\rm P}^{4h} \right)/2
	,\label{g1p}\\
	\partial_\ell g_{2\perp}&=
	g_{2\perp} g_{2\perp}
	\left(
	I_{\rm P}^{0} -I_{\rm C}^{0}
	\right)
	%
	-
	g_{1\perp} g_{1\perp}
	I_{\rm C}^{4h} ,
	\label{g2p}
	%
\end{align}
	%
	where the momentum dependences of the couplings on the left-hand sides of Eqs. (\ref{g||})-(\ref{g2p}) are $g_i (\ka, \kb, \kc)$ with incoming $\bk_1$, $-\bk_2$ and outgoing $\bk_3$ and $-\bk_4=\bk_1-\bk_2-\bk_3$.
	%
	%
	On the right-hand side, the momentum dependence  of  coupling products are in order: $g(\ka, \kp-\qcp,\kp) g(\kp, \kp-\qcp, \kc)$ for Cooper channel (particle-particle channel: SC fluctuations) that involves \hbox{$\IC^{\alpha h}(k_b,\qcp = \ka-\kb)$};  and $g(\ka, \Qp-\kp, \kp) g(\kp, \Qp-\kc, \kc)$ for the Peierls channel (particle-hole channel: DW fluctuations) that involves \hbox{$\IP^{\alpha h}(k_b,\Qp=\kb+\kc)$}. 
	The summations with respect to $\kp$ are taken as $N^{-1}\sum_{\kp}$, where $N$ is the number of chains in the $ab$ plane.
	The loop shell integration  yields
	\begin{align}
		I_\nu^{\alpha h} &=
		\frac{E_\ell}{4}
		\sum_{\lambda=\pm 1}
		\frac{\Theta (|E_\ell/2 +\lambda A_\nu^{\alpha h}|-E_\ell/2)}{E_\ell + \lambda A_\nu^{\alpha h}}
		\nonumber\\& \times
		\left\{
		\tanh \frac{E_\ell}{4T}
		+\tanh \frac{E_\ell/2 + \lambda A_\nu^{\alpha h}}{2T}
		\right\},
\end{align}
	where 
	\begin{align}
	A_{\rm C}^{\alpha h}
	&=\xi_b(\kp) - \xi_b(q_b-\kp)
	-\xi_b(\ka) + \xi_b(\kb)
	-\alpha h,\\
	A_{\rm P}^{\alpha h}
	&=\xi_b(\kp) + \xi_b(\kp-Q_b)
	-\xi_b(\kb) - \xi_b(\kc)
	-\alpha h,
	\end{align}
	and $\Theta (x) = 1$, $1/2$ and $0$ for $x>0$, $x=0$ and $x<0$, respectively.	
	The scaled bandwidth is  $E_\ell=E_0 e^{-\ell}$ at the RG step $\ell>0$.
	The initial band width $E_0 (=2v_F k_F)$ is taken as unity.
	The magnetic field acts as a cutoff for certain logarithmic singularities through $A_\nu^{\alpha h}$ in the Cooper and Peierls channels, in a way similar to nesting deviations $ t_b'$  for the Peierls singularity.
	
	From Refs. \cite{Bourbonnais2004,Fuseya2007a}, the dimensionless response function  for singlet type pairing at $\bq=0$, reads
\begin{align}
	\chi_{\rm ss}^{\rm e, o} (T)&=  \frac{2}{\pi v_{\rm F}}
	\int_0^\infty \!\!d\ell
	\left\{ z_{{\rm ss}, n}^{\rm e, o} (\ell)\right\}^2
	\IC^{2h} (\ell),
\end{align}
	where the pair vertex obeys the flow equation
\begin{align}
	\partial_\ell z_{{\rm ss}, n}^{\rm e, o}
	&=-\frac{1}{2}\langle 
	z_{{\rm ss}, n}^{\rm e, o} 
	(g_{1\perp}+g_{2\perp})
	\IC^{2h}
	\rangle _n^{\rm e, o}.
\end{align}
	The momentum dependences for the couplings and loop are $g_\mu \left(
	\kp, \kp-\qcp, \kp' \right)$ and $\IC^{2h}(q_b, \kp')$. 
	$\langle ... \rangle _n^{\rm e, o}$ denotes the even- (odd)-$n$-th Fourier component as $\langle f(k_b, k_b')\rangle_n^{\rm e(o)} = \langle f(k_b, k_b') \cos (\sin) nk_b\cos (\sin) nk_b'\rangle $.
	In particular, $\langle \cdots \rangle_0^{\rm e}$ is for $s$-wave and $\langle \cdots \rangle_1^{\rm e}$ is for  $d$-wave singlet, and $\langle \cdots \rangle_0^{\rm o}$ is for $p$-wave and $\langle \cdots \rangle_1^{\rm o}$ is for  $f$-wave triplet pairings\cite{Fuseya2005}.

	For the FFLO type pairing, $\bq \neq 0$,  and the formulation is modified due to the mixture of singlet and triplet pairing as follows:
	\begin{align}
	\chi_{{\rm ss}q}^{\rm e, o} (T)=  \frac{1}{\pi v_{\rm F}}
	\int_0^\infty \!\!d\ell
	&\bigl[\left\{ z_{{\rm ss}q, n}^{\uparrow \downarrow \rm e, o} (\ell)\right\}^2
	\IC^{0} (\ell)
	\nonumber\\&
	+\left\{ z_{{\rm ss}q, n}^{\downarrow \uparrow \rm e, o} (\ell)\right\}^2
	\IC^{4h} (\ell)
	\bigr],
\end{align}
	where the pair vertex is renormalized as
\begin{align}
	\partial_\ell z_{{\rm ss}q, n}^{\uparrow \downarrow \rm e, o}
	&=-\frac{1}{2}\langle 
	z_{{\rm ss}q, n}^{\uparrow \downarrow \rm e, o} g_{2\perp}\IC^0
	+z_{{\rm ss}q, n}^{\downarrow \uparrow \rm e, o} g_{1\perp}\IC^{4h}
	\rangle _n^{\rm e, o}.
\end{align}
	Similar definitions hold for the other pair response functions\cite{Fuseya2007a}.
	For the determination of the strongest singularity corresponding to the dominant instability, we compare the derivatives of the response functions, $\partial_\ell \chi_\mu$.
	In the following,  we set $t_b=0.1E_0$, $t_b'=0.01E_0$,   $g_{||}=0.0$ and finite repulsive $\tilde{U}$.

\section{Results}
\begin{figure}
\onefigure[width=7cm]{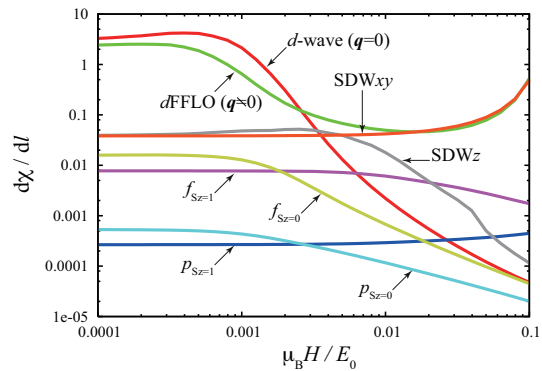}
\caption{Response functions, $\partial_\ell \chi_\mu (h)$, as a function of magnetic field for $\tilde{U}=1.0$, $T=10^{-3} E_0$ and $N=31$.
The subscript of $p$ and $f$ denotes the $s_z$ of triplet pairing; $s_z=1$ is the equal spin pairing and $s_z=1$ is the triplet pairing with antiparallel spins.
SDW$_{z}$ and SDW$_{xy}$ are the longitudinal- and the transverse spin density wave, respectively. }
\label{fig.1}
\end{figure}
	%
	{\it Response functions.}
	The magnetic field dependences of the derivative of the response functions, $\partial_\ell \chi_\mu (H)$, for $N=31$ chains are shown in Fig. \ref{fig.1}.
	Here we set $g_{1\perp}=g_{2\perp}=1.0$ ($\tilde{U}=1.0$), and $T=10^{-3} E_0$.
	The  SDW correlations are reduced (compared to $d$-wave) by  nesting deviations due to $t_{b}'$.
	At low fields, the most dominant instability is for $d$-wave pairing with $\bq=0$.
	(Hereafter we call the non-FFLO type ($\bq=0$) $d$-wave paring just as ``$d$-wave pairing".)
	It is reduced for $h\gtrsim 10^{-3} E_0 = T$, reflecting the cut-off effect of $h$.
	The instability  that closely follows is towards $d$-wave FFLO ($d$FFLO) order, which is also reduced for $h\gtrsim T$, albeit less than $d$-wave, so that it dominates at high field.
	In the limit of very high field, $h\gtrsim 10^{-2}E_0$, the $d$FFLO and the transverse SDW (SDW$_{xy}$) are  prevalent  with increasing $H$, 
	feeding one another through the coupling $g_{2\perp}$ as a result of interference. 
	%
	%
	%

\subsection{$H$-$T$ phase diagram}
\begin{figure}
\onefigure[width=7cm]{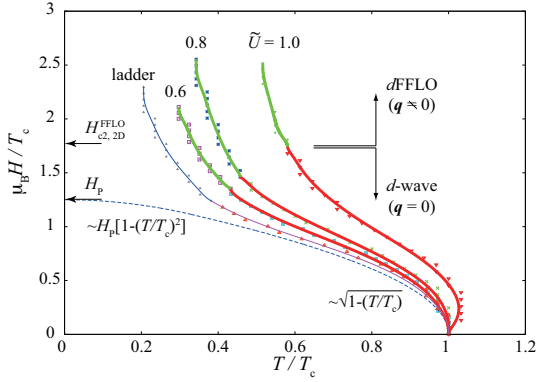}
\caption{$\Hc$ vs. temperature for $\tilde{U}=1.2, 1.0, 0.8$ (from top to bottom) for $N=15$.
The bottom curve is the $\Hc$ for  $\tilde{U}=-0.2$ without the Peierls channels, i.e., no SC-DW interference (ladder diagrammatic summation).
}
\label{fig.2}
\end{figure}
	
	Next, we show the results of the temperature dependence of $\Hc$ for $\tilde{U}=$0.6, 0.8, and 1.0 in Fig. \ref{fig.2}.
	Significant features are obtained.
	First the trace of $\Hc (T)$, which is determined by the singularity in $\partial_{\ell}\chi_{\mu}$, greatly exceeds not only the Pauli limit, $\HP \simeq 1.25 \Tc$, but also the paramagnetic limit of  the FFLO state, $H_{\rm c2}^{{\rm FFLO}}$ in 2D $(\simeq 1.78 \Tc$) and 3D ($\simeq 1.34 \Tc$ 
) isotropic cases\cite{Matsuda2007}.
	Significantly enough, the limiting field ratio $H_{\rm c2}^{d{\rm FFLO}}/T_c$ found here is non universal and depends on the strength of interactions.
	
	%
	%
	The origin of such an increase is primarily due to quasi-1D fluctuations introduced by SC-DW interference. 
	The mixing between the Peierls and Cooper scattering channels in Eq. (\ref{g1p}) induce an attractive coupling for $d$-wave superconductivity from spin fluctuations, which is scale dependent all the way down to the lowest energy scale of the flow, that is  $T_c$. 
	This contrasts  with  the  standard single channel or mean-field approach to superconductivity  for which the attractive coupling that feeds the Cooper pairing channel is fixed and tied to the high-energy cutoff of the theory. 
	This can be  further checked by comparing the RG results with those of the RPA --- ladder diagrammatic summation ---  limit, which is  obtained by switching off  the Peierls contribution  to the flow of couplings constants in Eqs. (2)-(4) ($I_P^{\alpha h}$=0). 
	There will be then  no interference and  by assuming an attractive coupling $\tilde{U}<0$,  an $s$-wave SC instability will thus  be realized in the mean-field way, while keeping the same geometric condition on the FS.
	
\begin{figure}
\begin{center}
\includegraphics[width=8.8cm]{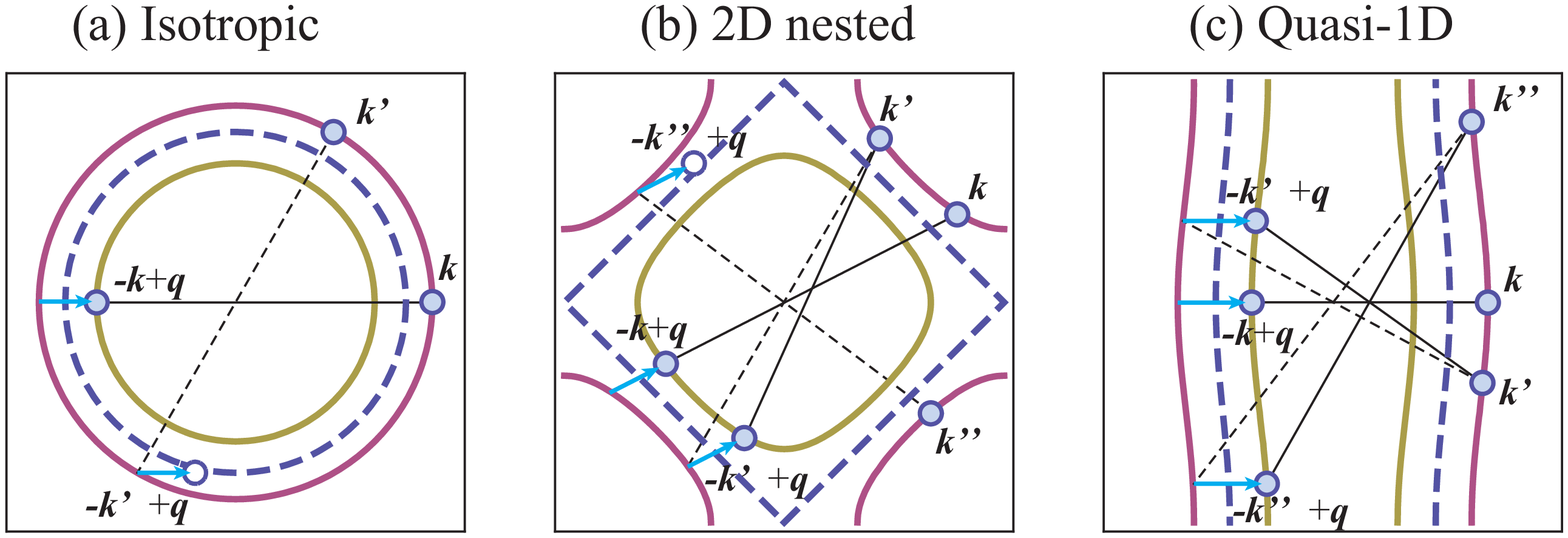}
\end{center}
\caption{ 
Possible FFLO pairings ($\bq\neq 0$) for (a) isotropic, (b) 2D nested, and (c) quasi-1D FS.
The FS at zero field is represented by the dashed lines. 
The pairs of $(\bk, -\bk+\bq)$ locate on the FS split by the Zeeman field.
(a) For the isotropic case, the pair of $(\bk', -\bk'+\bq)$ cannot locate on the FS; the formation of the FFLO pair is difficult.
(b) When the FS is nested, some of pairs like $(\bk', -\bk'+\bq)$  can locate on the FS, but others like $(\bk'', -\bk''+\bq)$ cannot. 
(c) In the case of the quasi-1D FS, almost all pairs can locate on the FS, so that there is no geometrical restriction.
}
\label{FS_FFLO}
\end{figure}

	The results of ladder summation  are also shown in Fig. \ref{fig.2} (the bottom solid curve).
	%
	%
	In this case $\Hc(T)$ almost agrees with the ordinary one\cite{Matsuda2007}, $\Hc(T)\simeq \HP \sqrt{1-T/\Tc}$ around $\Tc$, and extrapolates also towards $\HP$ at zero temperature.
	As for the FFLO state in the ladder limit, $\Hc^{\rm FFLO}(T)$ is higher than the isotropic value $H_{\rm c2, 2D}^{\rm FFLO}$.
	This slight increase of $\Hc^{\rm FFLO}(T)$ is due to a geometric property of the quasi-1D FS.
	In general, there is a geometric restriction to finite momentum  FFLO pairing, as  shown in Fig. \ref{FS_FFLO} (a).
	This restriction is partly removed when the FS  shows nesting  (Fig. \ref{FS_FFLO} (b)).
	In the case of  a quasi-1D FS, however, there is no restriction and  FFLO pairing is enhanced,  as   shown in Fig. \ref{FS_FFLO} (c). 
	In the ladder approximation, one finds that the traces for $\Hc/\Tc$ vs $T/T_c$, as obtained for different interactions $\tilde{U}$ (not shown), all scaled into the same   universal curve at variance with the nonuniversality displayed by the full one-loop RG results.
	%
	
	%
	%
	%
	%
	
	%
	%
	%

\section{Discussion}
\begin{figure}[h]
\onefigure[width=7cm]{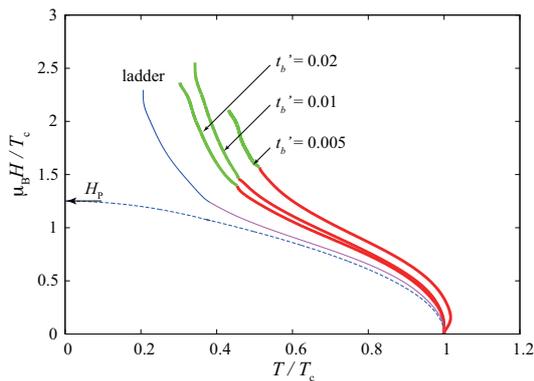}
\caption{Upper critical field for different anti-nesting paramters ($t_b' / E_0=0.005, 0.01, 0.02$) for $t_b /E_0 = 0.1$, $\tilde{U}=0.8$ and $N=15$.
The effect of $t_b'$ mimics the pressure effect in (TMTSF)$_2 X$.
}
\label{fig.5}
\end{figure}
	
	Let us now discuss the implications of the present results for  (TMTSF)$_2 X$ compounds.
	For (TMTSF)$_2$ClO$_4$, the specific heat measurements of bulk critical field, $\Hc^{\rm bulk}$,  along both $ a$ and $b'$ directions agree with the Pauli limit at low temperature, but it is significantly lower  than $\Hc^{\rm onset}$, as extracted from resistivity onset. 
	This allows to define a rather large ``transient region'', $\Hc^{\rm bulk} < H < \Hc^{\rm onset}$, in the $H$-$T$ plane where incomplete superconductivity is found.
	While there are no specific heat data available for (TMTSF)$_2$PF$_6$ under field, the values of $\Hc^{\rm onset}$ as obtained close to the critical pressure are similar to those of (TMTSF)$_2$ClO$_4$ at ambient pressure. 
	This suggests that the transient region for both compounds is of similar size. 

	The transient region observed experimentally should corresponds to the $d$FFLO region of the above two-dimensional theory (Fig. \ref{fig.2}).
	It seems to even exceed in size the one found experimentally at intermediate initial coupling $\tilde{U}$.
	%
	%
	However,  $d$FFLO superconductivity is expected to be quite sensitive to the finite impurity scattering in these materials, an effect known to more pronounced for $ H^{d{\rm FFLO}}_{c2}$\cite{Agterberg2001} than for $d$-wave critical $\Tc(H=0)$\cite{Joo2005}, which would result in some reduction of the transient region.
	Furthermore, the influence of orbital pair breaking effect, not included in the present Zeeman field theory but which is finite in practice, would also contribute to the reduction. 

	Finally, a crucial test of the present theory of $d$FFLO state in materials like the Bechgaard salts would be clearly the comparison of $H$-$T$ phase diagram at various  pressures, as it can be measured for instance by resistivity. 
	In the present model the spin fluctuations induced attractive coupling for $d$-wave superconductivity is known to be strongly pressure dependent via the anti-nesting parameter $t_b'$, which besides $\tilde{U}$ simulates  pressure effects in the model \cite{Nickel2006,Bourbonnais2009}.  
	Actually, the ratio $\Hc /\Tc$ decreases decreases with increasing $t_{b}'$ as shown in Fig. \ref{fig.5}, namely, the size of the transient region will shrink in size as pressure increase.
	The pressure dependent, non universal, reduction of $\Hc$ is a direct consequence of non universality linked to the interference between SDW and Cooper pairing. 	
	%
	%
	%
	%
	
	%
	%
	%
	%

\section{Summary}
	In the present work, $\Hc$ for the quasi-1D superconductivity have been calculated for the Hubbard model by the newly developed $N$-chain RG under magnetic field.
	We have shown that when constructive interference between density-wave and Cooper pairings is fully taken into account, the  $\Hc$ of a quasi-1D $d$-wave superconductor greatly exceeds not only the Pauli limit, but also the conventional paramagnetic limit of the FFLO state. 
	In a material like (TMTSF)$_2$ClO$_4$, the results presented here propose a new avenue  for the reconciliation of  transport\cite{Yonezawa2008a,Yonezawa2008b}, magneto-torque\cite{Satsukawa2009}, specific heat\cite{Yonezawa2012} and NMR data\cite{Shinagawa2007}.  
	The anomalous $\Hc$ enhancement found from this mechanism proves to be non universal in both interaction and anti-nesting strengths, allowing  the possibility for a direct experimental test for the predicted correlation between  the $H_{c2}$-excess and $T_c$ under pressure.
	%
	%
	
	In conclusion, we have proposed a mechanism of the anomalous broadening of the upper critical field region as seen from transport measurements in quasi-1D organic superconductors like  the Bechgaard salts series. 
	The mechanism  proposed is a direct consequence of the interplay between magnetism and unconventional superconductivity in these materials under magnetic field. 
	%
	%
	%
	%


\acknowledgments
	%
	We gratefully acknowledge S. Yonezawa for discussions and providing the experimental data.
	We are also grateful to D. J\'erome, Y. Suzumura, M. Ogata and M. Tsuchiizu for fruitful discussions.
	One of the authors (Y. F.) is supported by a Grant-in-Aid for Young Scientists (Start-up) from JSPS, and JSPS Research Fellowships for Young Scientists.

\end{document}